\documentclass[aps,pre,reprint,amsmath,amssymb]{revtex4-2}

\usepackage{graphicx}
\usepackage{txfonts}
\usepackage{url}
\usepackage{color}
\usepackage{enumitem}
\usepackage{bm}
\usepackage{algorithm}
\usepackage{algorithmic}
\usepackage{here}

\begin{document}

\title{Koopman analysis of combinatorial optimization problems with replica exchange Monte Carlo method}

\author{Tatsuya Naoi, Tatsuya Kishimoto, and Jun Ohkubo}

\affiliation{Graduate School of Science and Engineering, Saitama University, Sakura, Saitama 338--8570 Japan}

\begin{abstract}
Combinatorial optimization problems play crucial roles in real-world applications, and many studies from a physics perspective have contributed to specialized hardware for high-speed computation. However, some combinatorial optimization problems are easy to solve, and others are not. Hence, the qualification of the difficulty in problem-solving will be beneficial. In this paper, we employ the Koopman analysis for multiple time-series data from the replica exchange Monte Carlo method. After proposing a quantity that aggregates the information of the multiple time-series data, we performed numerical experiments. The results indicate a negative correlation between the proposed quantity and the ability of the solution search.
\end{abstract}

\maketitle

\section{Introduction}
\label{sec_introduction}

Combinatorial optimization problems play a crucial role in various fields in real-world applications, and their solutions contribute to efficient resource utilization, cost reduction, and service quality improvement. Examples include supply chain management, finance, manufacturing, transportation, energy, healthcare, and telecommunications. In combinatorial optimization problems, we find combinations of variables that minimize an objective function while satisfying various constraints. Since the minimization procedure is the same as seeking ground energy states in statistical physics \cite{Kirkpatrick1983}, there are many works for combinatorial optimization problems from a physics viewpoint. For example, D-wave Advantege by D-Wave systems \cite{Bunyk2014,d-wave,d-wave-company} employs the principle of quantum annealing \cite{Kadowaki1998,Farhi2001}. Another type of hardware is based on complementary metal-oxide semiconductors (CMOS), including devices developed by Hitachi \cite{Takemoto2019,Takemoto2021}, Toshiba \cite{Goto2019a,Goto2019b,Tatsumura2019}, and Fujitsu \cite{Aramon2019,Matsubara2020}. In the CMOS-type hardware, an architecture based on the replica exchange Monte Carlo method \cite{Swendsen1986,Hukushima1996} is sometimes employed, which exploits the merit of parallel computation.

While recent annealing hardware enables us to solve combinatorial optimization problems at high speed, the time required to find the optimal solution varies from problem to problem. For example, a class of 0-1 quadratic knapsack problem \cite{Gallo1980} is NP-hard, but some are easy to solve, and others are not. Conventional simulated or quantum annealing requires a slow reduction in temperature or quantum effects when the combinatorial optimization problem is hard to solve. Although there is a well-known schedule that yields an optimal solution \cite{Geman1984,Suzuki2005}, the schedule is too late in practice. In addition, it is worth knowing in advance whether the problem is easy to solve or not.

Here, we expect that the time-series data used to find the solution will reflect the degree of difficulty in problem-solving; the dynamics for easy problems will have simple structures, while difficult problems yield complicated dynamics. To investigate the dynamics, one can employ the Koopman theory \cite{Koopman1931}, which has attracted attention in various fields. The Koopman theory enables us to use linear algebra for nonlinear systems; it is also possible to analyze stochastic nonlinear systems within the framework of linear algebra. In the Koopman theory, we deal with a Koopman operator for the time evolution in the observable function space, not the state space. While the observable function space is infinite-dimensional, the Koopman operator is linear even if the underlying dynamical system is nonlinear. Several methods have been developed in recent years to obtain Koopman operators from an observed data set. One of the methods is dynamic mode decomposition (DMD), which analyzes dynamical systems from observed data \cite{Rowley2009}. There are extended methods, such as extended dynamic mode decomposition (EDMD) \cite{Williams2015}. The EDMD can efficiently approximate the Koopman operator as a finite-dimensional matrix. Applications of EDMD are currently being studied, including time series prediction, system identification \cite{Mauroy2020}, and control \cite{Korda2018}.

In this paper, we aim to clarify the relationship between the Koopman analysis and the degree of difficulty in combinatorial optimization problems. The EDMD yields an approximated matrix, the so-called Koopman matrix, for the Koopman operator. Since the replica exchange Monte Carlo method generates time-series data for each temperature, we get several datasets for the Koopman analysis. After applying the EDMD to time-series data generated from the replica exchange Monte Carlo method, we analyze the eigenvalues of the derived Koopman matrix for each temperature; the information of the eigenvalues will contain that of the degree of difficulty for each problem. We will also discuss a conventional data analysis based on singular value decomposition (SVD).

This paper is organized as follows. Section~\ref{sec_background} reviews the background of the annealing method and the Koopman theory. Section~\ref{sec_proposal} yields the main proposal, including data acquisition and analysis methods. In Sect.~\ref{sec_experiment}, we demonstrate the proposed method with numerical experiments on randomly generated problems and the 0-1 quadratic knapsack problems; the comparisons with the proposed method and the conventional SVD are also discussed. We give a summary and mention of future work in Sect.~\ref{sec_conclusion}.

\section{Background}
\label{sec_background}

\subsection{QUBO formulation and Ising model}

A quadratic form of binary variables called quadratic unconstrained binary optimization (QUBO) formulation is widely used as an input for specialized hardware. The QUBO formulation is deeply related to the Ising model, and both have binary variables. The QUBO formulation has the following cost function for the state vector $\bm{a}$:
\begin{align}
\label{func_qubo}
E(\bm{a}) = \sum_{i\in\mathcal{D}}\sum_{j\in\mathcal{D}} \frac{1}{2} Q_{ij}a_{i}a_{j},
\end{align}
where $\mathcal{D}$ is the set of indices of the variables, $a_{i} \in \{0,1\}$ is the $i$-th binary variable in $\bm{a}$, and $Q_{ij} \in \mathbb{R}$ is the strength of the interaction between the binary variables $a_{i}$ and $a_{j}$. The number of spins is $N$, i.e.,  $|\mathcal{D}| = N$. By contrast, the Ising model has binary spins with $\sigma_{i} \in \{-1,1\}$, and the energy (cost) function of the Ising model is denoted as follows:
\begin{align}
\label{func_ising}
E(\bm{\sigma}) = -\sum_{i\in\mathcal{D}}\sum_{j\in\mathcal{D}} \frac{1}{2} J_{ij}\sigma_{i}\sigma_{j}-\sum_{i\in\mathcal{D}}h_{i}\sigma_{i},
\end{align}
where $\bm{\sigma}$ is the spin vector, $J_{ij}\in{\mathbb{R}}$ is the two-body interaction between the spins $\sigma_{i}$ and $\sigma_{j}$, and $h_{i}$ is the external magnetic field on $\sigma_{i}$.

The energy function of the Ising model is equivalent to the cost function of QUBO formulation via the following transformation:
\begin{align}
\label{transform}
a_{i} = \frac{1+\sigma_{i}}{2}.
\end{align}
Conversion between $\{J_{ij}\},\{h_{i}\}$ and $\{Q_{ij}\}$ is also possible.

It is possible to convert various combinatorial optimization problems into the QUBO formulation \cite{Lucas2014}. Hence, minimizing the functions in Eqs.~\eqref{func_qubo} or \eqref{func_ising} corresponds to seeking solutions to combinatorial optimization problems.

\subsection{Replica exchange Monte Carlo method}

The replica exchange Monte Carlo method, also known as the parallel tempering method, improves sampling efficiency in Monte Carlo simulations and Markov chain Monte Calro methods \cite{Swendsen1986,Hukushima1996}. In the replica exchange method, the temperature of each replica is determined and fixed in advance, and each replica develops with the conventional Metropolis rule. There is an exchange procedure for replicas. Due to the replica exchanges, when a low-temperature replica gets stuck in a local minimum, the replica can escape from the local minimum by exchanging a higher-temperature replica. 

Let $R$ be the number of replicas and $\bm{a}^{(r)}$ be the state vector for the $r$-th replica. Then, the Gibbs distribution is defined as follows:
\begin{align}
P(\bm{a}^{(r)})=\frac{1}{Z^{(r)}}\exp\left(-\frac{E(\bm{a}^{(r)})}{T^{(r)}}\right),
\end{align}
where $T^{(r)}$, $E(\bm{a}^{(r)})$, and $Z^{(r)}$ correspond to the temperature parameter, the cost function, and the normalization constant for the $r$-th replica, respectively. While it is difficult to evaluate the normalization constant $Z^{(r)}$ in general, there is no need to evaluate it; the state transition obeys the following conventional Metropolis rule:
\begin{align}
P^{(r)}_\mathrm{change}=\mathrm{min}\left[1,\exp\left(-\frac{\Delta E^{(r)}}{T_{r}}\right)\right],
\label{eq_Metropolis}
\end{align}
where $\Delta E^{(r)}$ is the energy difference from the previous state to the next one in the $r$-th replica. The Metropolis rule yields a sampling from the Gibbs distribution without evaluating the normalization constant $Z^{(r)}$.

As denoted above, there is the exchange procedure of the state vectors between different replicas. Although low-temperature settings are necessary when seeking stable states, there are many local minima that have difficulty escaping at low temperatures. Hence, we seek various configurations using the replicas with high temperatures. Here, we define the temperature parameters in ascending order: $T_{1} < T_{2} < \cdots < T_{R}$. Then, the probability $P^{(m,l)}_\mathrm{exchange}$ for the exchange between the $m$-th and $l$-th replicas is defined as
\begin{align}
\label{exchange_replica}
P^{(m,l)}_\mathrm{exchange}=\mathrm{min}\left[1,\exp\left\{\left(E^{(m)}-E^{(l)}\right)\left(\frac{1}{T_{m}}-\frac{1}{T_{l}}\right)\right\}\right],
\end{align}
where $E^{(m)}$ and $E^{(l)}$ are the energies of the $m$-th and $l$-th replicas, respectively. In this paper, we consider only exchanges at two adjacent replicas, and the following equation holds in Eq.~\eqref{exchange_replica}:
\begin{align}
l = m + 1
\end{align}
for $m=1, \cdots, R-1$.

Note that the data analysis method proposed in Sect.~\ref{sec_proposal} assumes the usage of the replica-exchange Monte Carlo method. As denoted in Sect.~\ref{sec_introduction}, the replica-exchange Monte Carlo method is employed on a certain type of annealing hardware and is also suitable for parallel computing. Hence, the assumption of using the replica-exchange Monte Carlo method is appropriate.

\subsection{Koopman theory}

We here briefly review the Koopman theory; for more details, see, for example, the review paper in Ref.~\cite{Brunton2022}.

The Koopman operator is a linear operator defined mainly in nonlinear dynamical systems. In the Koopman theory, we consider the time evolution of an observable function rather than that of the state variables. Consider the following time-evolution:
\begin{align}
\label{states}
\bm{a}_{t+1} = F(\bm{a}_{t}),
\end{align}
where $\bm{a}_{t}$ is a state vector at time $t$, and $F$ is a nonlinear time evolution operator. Here, we consider a deterministic dynamics for simplicity; a stochastic case will be commented later. We also introduce an observable function $\phi(\bm{a})$. For example, $\phi(\bm{a}) = a_i$ corresponds to the observation of the $i$-th spin variable state. Then, the Koopman operator $\mathcal{K}$ acts on the function $\phi$ as follows:
\begin{align}
\label{observables}
\mathcal{K}\phi = \phi \circ F.
\end{align}
Equations~\eqref{states} and \eqref{observables} lead to
\begin{align}
\label{Koopmanlinear}
\phi(\bm{a}_{t+1})=(\mathcal{K}\phi)(\bm{a}_{t}).
\end{align}
It is easy to confirm the linearity of the Koopman operator; for any constants $c_1, c_2 \in \mathbb{R}$ and observable functions $\phi_1, \phi_2$, we have
\begin{align}
\left\{ \mathcal{K}(c_1 \phi_1 + c_2 \phi_2) \right\} (\bm{a}_t) &= (c_1 \phi_1+c_2 \phi_2)(\bm{a}_{t+1}) \nonumber\\
&= c_1 \phi_1(\bm{a}_{t+1}) + c_2 \phi_2(\bm{a}_{t+1}) \nonumber\\
&= c_1(\mathcal{K}\phi_1)(\bm{a}_t) + c_2(\mathcal{K}\phi_2)(\bm{a}_t) \nonumber\\
&= (c_1\mathcal{K}\phi_1 + c_2\mathcal{K}\phi_2)(\bm{a}_t).
\end{align}

Note that the Koopman operator $\mathcal{K}$ is infinite-dimensional because $\mathcal{K}$ acts on elements in a function space.  Therefore, it is necessary to approximate the Koopman operator $\mathcal{K}$ as a finite-dimensional Koopman matrix $K$ in practice. To describe the Koopman matrix $K$, we introduce a so-called dictionary. The dictionary is defined as
\begin{align}
\label{dictionary}
\bm{\psi}(\bm{a}) = [\psi_{1}(\bm{a}),\psi_{2}(\bm{a}),\cdots,\psi_{N_\mathrm{dic}}(\bm{a})]^{\top},
\end{align}
where $\psi_{i}(\bm{a})$ is the $i$-th function and $N_\mathrm{dic}$ is the size of the dictionary. Note that in Ref.~\cite{Williams2015}, the dictionary is defined as a row vector, but in this paper we consider it as a column vector. Then, we introduce the following linear combination to express an observable function:
\begin{align}
\label{eq_phi_linear_combination}
\phi(\bm{a}_{t}) = \sum_{k=1}^{N_\mathrm{dic}}c_{k}\psi_{k}(\bm{a}_{t}) = \bm{c}^{\top}\bm{\psi}(\bm{a}_{t}),
\end{align}
where $\bm{c}$ is a coefficient vector. Combining Eqs.~\eqref{Koopmanlinear} and \eqref{eq_phi_linear_combination}, we have
\begin{align}
\left(\mathcal{K}\phi\right)(\bm{a}_{t})  = \bm{c}^{\top} \left(\mathcal{K}\bm{\psi}\right)(\bm{a}_{t}).
\end{align}
Note that $\bm{c}$ is time-independent. Hence, instead of the action of the Koopman operator on the observable function $\phi$, it is enough to consider the action on the dictionary as follows:
\begin{align}
\bm{\psi}(\bm{a}_{t+1}) = \mathcal{K} \bm{\psi}(\bm{a}_{t}) \simeq K \bm{\psi}(\bm{a}_{t}),
\label{eq_Koopman_matrix}
\end{align}
which leads to the Koopman matrix $K$.

Here, we comment on the dimensionality of the Koopman operator $\mathcal{K}$. As denoted above, the dimension of the function space is infinite. We can understand this fact from a Taylor expansion of a function; the expansion consists of a polynomial with an infinite number of terms. However, for the QUBO cases, it is enough to consider a subspace in the infinite-dimensional space. Since each variable $a_i$ only takes $0$ or $1$, $a_i^2$ equals $a_i$. Hence, only $2^N$ terms remain. Of course, the dimensionality is still high, and some approximation is necessary. We will introduce an explicit dictionary in Sect.~III.B.

The EDMD is a method to derive the Koopman matrix $K$ from a dataset \cite{Williams2015}. Here, we consider a single time-series dataset $\{\bm{a}_{1},\bm{a}_2,\dots,\bm{a}_{M+1}\}$, although it is sufficient to have snapshot pairs rather than the single time-series data. Then, the least-squares problem with the cost function,
\begin{align}
J = \sum_{t=1}^M \big\| \bm{\psi}(\bm{a}_{t+1}) - K\bm{\psi}(\bm{a}_{t}) \big\|^2,
\end{align}
leads to the Koopman matrix $K$ immediately. The solution is denoted as
\begin{align}
\label{matrixK}
K = QG^{+},
\end{align}
where 
\begin{align}
G = \frac{1}{M}\sum_{t=1}^{M}\bm{\psi}(\bm{a}_{t})^{\top}\bm{\psi}(\bm{a}_{t}), \quad
Q = \frac{1}{M}\sum_{t=1}^{M}\bm{\psi}(\bm{a}_{t+1})^{\top}\bm{\psi}(\bm{a}_a{t}),
\end{align}
and $G^{+}$ is the pseudo-inverse of the matrix $G$.

In this paper, the Metropolis rule in Eq.~\eqref{eq_Metropolis} yields the dynamics, and hence we should consider a stochastic dynamics. As discussed in Ref.~\cite{Williams2015}, the Koopman matrix $K$ gives expectations in the stochastic case; Eq.~\eqref{eq_Koopman_matrix} is replaced simply with
\begin{align}
\mathbb{E}\left[\bm{\psi}(\bm{a}_{t+1})\right] \simeq K \bm{\psi}(\bm{a}_{t}).
\end{align}
Even in the stochastic case, the Koopman matrix $K$ contains the information of the system dynamics.

\section{Difficulty in Problem-solving and Koopman Analysis}
\label{sec_proposal}

Here, we will discuss a relationship between the Koopman analysis and the degree of difficulty in problem-solving. 

\subsection{Flow for analyzing data}

One would expect that the dynamics for easy problems will have simple structures, while difficult problems yield complicated dynamics. The Koopman matrix contains the degree of complexity of the dynamics; it is possible to describe the dynamics of simple structures with a small number of modes, and the dynamics of complex structures will require many modes. Therefore, the distribution of eigenvalues should reflect the information of the dynamics.

However, a single time series of data alone cannot determine whether the eigenvalue distribution is simple or not. For example, the state of any problem will hardly change at considerably low temperatures, and only a few modes will remain. Therefore, we focus on the variation of the eigenvalue distributions at different temperatures. If the difficulty in problem-solving is high, the number of modes needed to describe the dynamics will increase as soon as the temperature becomes high.

\begin{figure}[bt]
  \centering
  \includegraphics[width=85mm]{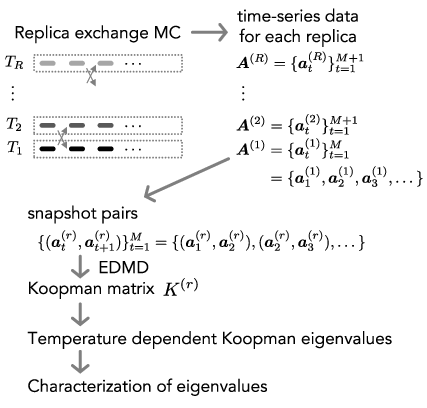}
  \caption{Flow for analyzing data. Finally, we obtain several Koopman matrices and their eigenvalues. The characteristics of the Koopman eigenvalue distribution are summarized as a simple summation of them; see Fig.~\ref{fig_proposed_2} and its explanation.}
  \label{fig_proposed_1}
\end{figure}

From the above considerations, we employ several time-series datasets generated from the replica exchange Monte Carlo method. Figure~\ref{fig_proposed_1} shows the flow for analyzing data. The replica exchange Monte Carlo method generates several time-series data with different temperature parameters, and we simply analyze each of those datasets. Although the flow is simple, there are some notices for the data analysis. We explain each of those notices below.

\subsection{Reduction of dictionary size}

The dictionary size in the EDMD becomes enormously large when the number of variables increases. For simplicity, consider monomial dictionary functions with the form $a_1^{p_1} a_2^{p_2}\cdots a_N^{p_N}$ where $p_i \in \mathbb{N}_0$. When $L-1$ is the maximum degree of each variable, i.e., $p_i < L$ for all $i$, the dictionary size is $L^N$. Of course, we can restrict $L$ to be less than two because the spin only takes $0$ or $1$; for example, $a_1^2 = a_1$. However, we still have the dictionary size with $\mathcal{O}(2^N)$. Due to the curse of dimensionality, we cannot evaluate the Koopman matrix.

In the data analysis, we restrict the total degree of monomials to be less than two and remove $a_i^2$ for $i = 1, \dots,N$ from the dictionary. Hence, we employ the following dictionary:
\begin{align}
\bm{\psi}(\bm{a}) =
\left[
1, a_1, a_2, \dots, a_N, a_1 a_2, a_1 a_3, \dots, a_{N-1} a_{N}
\right]^{\top}.
\label{eq_proposed_dictionary}
\end{align}
Preliminary numerical verification confirms that a higher-order dictionary, which also includes monomials of degree 3 or higher, does not significantly affect the results.  We consider the reason for this is that the QUBO formulation has only up to two-body interactions. Hence, we employ the dictionary in Eq.~\eqref{eq_proposed_dictionary} in the data analysis.

\subsection{Preparation of snapshot pairs}

As denoted in Fig.~\ref{fig_proposed_1}, there are replica exchanges between different temperature settings. In the EDMD, the analysis is based on snapshot pairs. Hence, one might suspect that the snapshot pairs at replica exchange timings could cause some problems in the data analysis; the snapshot pairs at the timings do not reflect the true dynamics of the system. 

For the preparation of the snapshot pairs, we performed preliminary numerical experiments in which we removed the snapshot pairs at the replica exchange timings. The numerical results showed no significant differences due to data preprocessing. The reason is the small number of replica exchanges; these snapshot pairs do not affect the data analysis. Hence, in the flow in Fig.~\ref{fig_proposed_1}, we do not remove the snapshot pairs at the replica exchange timings for simplicity.

\subsection{Summation of eigenvalues}

After the flow in Fig.~\ref{fig_proposed_1}, we have several eigenvalue distributions. As denoted above, we investigate the variation of the eigenvalue distribution at different temperatures.

\begin{figure}[tb]
  \centering
  \includegraphics[width=75mm]{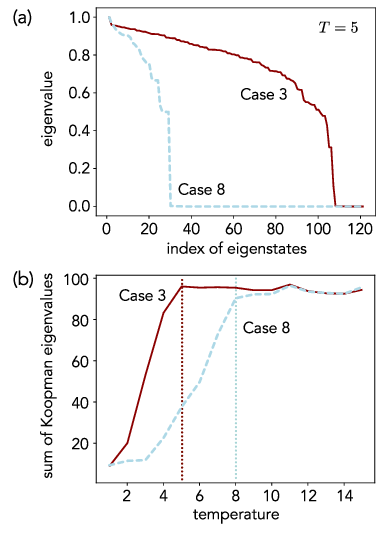}
  \caption{(Color online) (a) Examples of eigenvalue distributions. The case numbers correspond to those in Fig.~\ref{fig_random_Koopman}. (b) Temperature dependency of the total summation of eigenvalues. The left side of the dotted line in (b) is the region showing sharp temperature dependency for each case.}
  \label{fig_proposed_2}
\end{figure}

Figure~\ref{fig_proposed_2}(a) shows examples of the eigenvalue distributions; the experimental setting is the same with Sect.~\ref{subsec_random_problem}; we will explain it later. In Fig.~\ref{fig_proposed_2}(a), the distribution varies depending on the problem. There are several ways to capture the characteristics of the eigenvalue distributions. For example, the number of eigenvalues above a certain threshold would represent the number of dominant modes. After trying some ways, we finally decided to employ the summation of the eigenvalues. Figure~\ref{fig_proposed_2}(b) plots the sum of eigenvalues versus each temperature; we see clear differences depending on the problem.

\subsection{Characterization of temperature dependency}
\label{subsec_characteristics}

Finally, we characterize the shape of the temperature dependency in Fig.~\ref{fig_proposed_2}(b). After some trials, we found that the area of the rapidly changing region of the temperature dependency in Fig.~\ref{fig_proposed_2}(b) captures the characteristics of the difficulty of problem-solving. For example, the left side of the dotted line in Fig.~\ref{fig_proposed_2}(b) corresponds to the region showing sharp temperature dependency for each case.

Let $w_{r}$ be the sum of Koopman eigenvalues at the $r$-th temperature. We introduce the following quantity $S$ to characterize the difficulty in problem-solving:
\begin{align}
\label{area}
S = \sum^{R'}_{r=1}\left(\frac{1}{2}\left(w_{r+1}+w_{r}\right)\left(T_{r+1}-T_{r}\right)\right),
\end{align}
where $T_r$ is the temperature of the $r$-th replica, and $R'$ is the index of the end temperature of the area. Although one can determine the index $R'$ by the figure appearance, we here employ the following determination method:
\begin{enumerate}
  \item Set $r = 3$.
  \item Calculate $\Delta_r = (w_{r+1}-w_{r})/(w_{r}-w_{r-1})$.
  \item If $\Delta_r < 0.5$, then $R'=r$. If not, $r \rightarrow r+1$ and go back to Step 2.
\end{enumerate}
This procedure does not use the first few indices because the data does not show a monotonic increase in the first three or so temperature ranges. Additionally, this rule is not universally applicable; increasing the number of replicas and sampling more finely across temperature ranges could not necessarily exhibit a monotonic increase. However, in the numerical experiments below, we determine $R'$ in the manner described above to automate the procedure.

\section{Numerical Experiment}
\label{sec_experiment}

We perform several numerical experiments to check whether the proposed method successfully captures the characteristics of the difficulty of problem-solving. In the following numerical experiments, we deal with two examples; one is randomly generated, and the other is the 0-1 quadratic knapsack problem \cite{Gallo1980}, which was also used in the previous work for the specialized hardware \cite{Yin2023}. As for the 0-1 knapsack cases, we generate some problems randomly to vary the difficulty level of problem-solving.

The purpose of the following numerical experiments is to examine the relationship between the Koopman analysis and difficulty in problem-solving. Hence, we compare the number of finding times of the optimal solution with the quantity proposed in Sect.~\ref{subsec_characteristics}. Note that we consider only problems with small spin numbers because we need optimal solutions for the comparison.

\subsection{Randomly generated problems}
\label{subsec_random_problem}

Although we tried several problem settings, the following method for generating problems of varying difficulty is employed here. Here, the number of spins is $15$, and we randomly generate $J_{ij}$ in Eq.~\eqref{func_ising} from $\mathcal{N}(3,0.5)$ or $\mathcal{N}(-3,0.5)$ with equal probability. The coefficients $h_{i}$ in Eq.~\eqref{func_ising} are also generated from $\mathcal{N}(2,0.5)$ or $\mathcal{N}(-2,0.5)$ with equal probability. The number of replica is $R=15$, and the temperatures $\{T_{r}\}$ for $r=1, 2, \cdots, R$ are defined as follows:
\begin{align}
\label{temp_ising}
T_{r} = r + 1.0.
\end{align}
We obtain $10,000$ time-series data for each temperature by using the replica exchange Monte Carlo method. Before the analysis by the EDMD, the binary variables $\{-1,1\}$ are converted to $\{0,1\}$ by Eq.~\eqref{transform}. Here, we generate nine different problems, and for each problem, the quantity $S$ by the data analysis flow described in Sect.~\ref{sec_proposal} is evaluated.

\begin{figure}[tb]
  \centering
  \includegraphics[width=75mm]{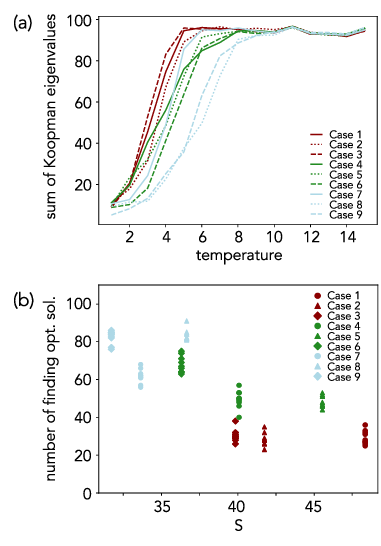}
  \caption{(Color online) Numerical results of the Koopman analysis for randomly generated problems. (a) shows the temperature-dependency of the total eigenvalues. (b) shows the relationship between the number of finding times of the optimal solution in 100 trials and the feature $S$.}
  \label{fig_random_Koopman}
\end{figure}

Figure~\ref{fig_random_Koopman}(a) shows the temperature-dependency of the sum of Koopman eigenvalues. For the evaluation of the quantity $S$ in Eq.~\eqref{area}, $R'$ is determined by the procedure described in Sect.~\ref{subsec_characteristics} as follows: 
\begin{itemize}
  \item $R'=5$ for Cases 1, 2, 3, 4 and 7,
  \item $R'=6$ for Cases 5 and 6,
  \item $R'=7$ for Case 9,
  \item $R'=8$ for Case 8.
\end{itemize}
As depicted in Fig.~\ref{fig_random_Koopman}(a), the sum of Koopman eigenvalues shows various behaviors in a low-temperature region. By contrast, at a high-temperature region, almost random transitions occur regardless of the problem's difficulty. Hence, the snapshot pairs show similar behavior, which causes the temperature-independent features.

Next, we measure the number of finding times of the optimal solution for each case. For the $15$ spins problem, the number of states is $2^{15} = 32,768$, and hence, it is possible to search exhaustively to find the state that minimizes the cost function. Note that the temperature parameters used to solve the problem in the replica exchange Monte Carlo method differ from those used to compute the features $S$; we employ the following conventional settings:
\begin{align}
  T_{r} = 0.001 + \left(\frac{r}{R}\right)^{2}, \quad (R=5 \textrm{ and } r=1, 2, \cdots, R).
\label{eq_temperature_setting_for_search}
\end{align}
We apply the updates for a randomly chosen spin $N$ times and the replica exchange procedure at the timing; the replica exchange occurs between a randomly selected replica number and the number one above it. We call this procedure an iteration step. After repeating the iteration steps 100 times, we judge whether we find the optimal solution. We repeat the search 100 times and count the number of finding times of the optimal solution. Figure~\ref{fig_random_Koopman}(b) shows the relationship between the number of finding times of the optimal solution in 100 trials and the feature $S$; it shows the results of ten runs of the above experiment for each problem case.

In Fig.~\ref{fig_random_Koopman}(b), we see a clear negative correlation between the number of finding times the optimal solution and the feature $S$. We note that the choice of the temperature range in evaluating Eq.~\eqref{area} has an arbitrariness, and the choice yields slightly different results. However, as far as we checked, the negative correlation still appeared despite the slightly different temperature ranges. Hence, these numerical results indicate that our conjecture is valid: the temperature-dependency in the sum of Koopman eigenvalues captures the difficulty in problem-solving.

\subsection{Quadratic knapsack Problem}
\label{subsec_QKP}

The 0-1 quadratic knapsack problem extends the standard knapsack problem. The standard knapsack problem aims to select items within a given capacity constraint to maximize the total value. However, in the 0-1 quadratic knapsack problem, the value of the items is represented in a quadratic form. As denoted above, the 0-1 quadratic knapsack problem was discussed in the specialized hardware previously \cite{Yin2023}. The cost function is denoted as follows:
\begin{align}
\label{func_qkp}
E(\bm{x}) = -\frac{1}{2}\sum_{i\in D}\sum_{j\in D}Val_{ij}x_{i}x_{j} - \sum_{i\in D}Val_{i}x_{i} \notag \\
+\lambda \mathrm{max}\left(0,\sum_{i\in D}WT_{i}x_{i}-WT_\mathrm{max}\right),
\end{align}
where $Val_{i}$ and $WT_{i}$ mean the value and the weight of the $i$-th item, respectively; $Val_{ij}$ is the interaction value between the $i$-th item and $j$-th item, and $WT_\mathrm{max}$ is the capacity of the knapsack. Note that $x_{i} \in \{0,1\}$.

Here, we consider problems with $15$ items. The problem parameters,  $Val_{ij}, Val_{i}, WT_{i},$ and $WT_\mathrm{max}$ in Eq~\eqref{func_qkp} are randomly generated from uniform distribution as follows:
\begin{itemize}
  \item $Val_{ij}\sim U(1.0,30.0) \,\, (i\neq j)$,
  \item $Val_{i}\sim U(1.0,30.0)$, 
  \item $WT_{i}\sim U(1.0,30.0)$, 
  \item $WT_\mathrm{max} \sim U^{\mathrm{integer}}(15,16,\cdots,300)$,
\end{itemize}
where $U^{\mathrm{integer}}$ has the integer domain. For $i=j$, $Val_{ij}=0$. The penalty constant $\lambda$ in Eq.~\eqref{func_qkp} is $100$. Note that the cost function is not the QUBO formulation. There are a few ways to deal with the penalty terms for the inequality constraints. For example, we can convert the penalty terms into the QUBO formulation via duality relations \cite{Sato2019}. In Ref.~\cite{Yin2023}, several techniques, such as asymmetric two-body interactions and slack variables, were employed. However, we here directly evaluate the cost function value from Eq.~\eqref{func_qkp} for simplicity because this paper aims to check the ability of the Koopman theory. Hence, the spin number is the same as that of items; there are $15$ spins.

For the evaluation of $S$, we use the following temperature settings:
\begin{align}
\label{temp_qkp}
T_{r} = 10 \times r + 10,
\end{align}
and $R = 10$. Other settings for the time-series data analysis are the same as in Sect.~\ref{subsec_random_problem}. In the solution search procedure, we use the same temperature parameters with Eq.~\eqref{eq_temperature_setting_for_search}. The number of iterations for each search is $200$, and the other settings for the search procedure are the same as in Sect.~\ref{subsec_random_problem}.

\begin{figure}[tb]
  \centering
  \includegraphics[width=75mm]{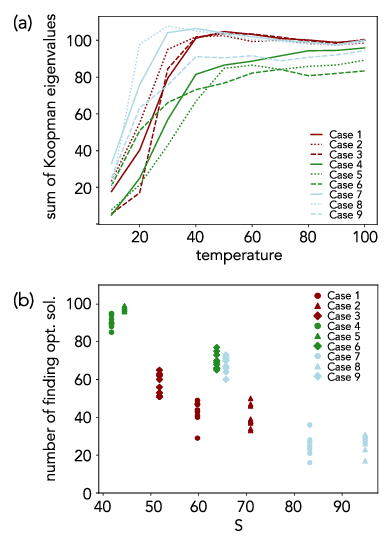}
  \caption{(Color online) Numerical results of the Koopman analysis for the 0-1 quadratic knapsack problems. (a) shows the temperature-dependency of the total eigenvalues. (b) shows the relationship between the number of finding times of the optimal solution in 100 trials and the feature $S$.}
  \label{fig_qkp_Koopman}
\end{figure}

Figure~\ref{fig_qkp_Koopman}(a) shows the temperature-dependency of the sum of Koopman eigenvalues. The values of $R'$ are as follows:
\begin{itemize}
\item $R'=4$ for Cases 1, 2, 3, 4, 7, and 9,
\item $R'=5$ for Case 5,
\item $R'=6$ for Cases 6 and 8.
\end{itemize}
Figure~\ref{fig_qkp_Koopman}(b) shows the relationship between the number of finding times of the optimal solution in 100 trials and the feature $S$; as in Sect.~\ref{subsec_random_problem}, there is a clear negative correlation even in the 0-1 knapsack problem.

\subsection{Data analysis with SVD}

The SVD is well-known for capturing the characteristics of data matrices. Therefore, we here analyze the time-series data with the SVD. When there are $N$ spins and the length of the time-series data is $M$, the time-series data yields a $N \times M$ data matrix. We construct the data matrix of the time-series data for each temperature and apply the SVD to the constructed data matrices. In the EDMD method, we investigated the temperature dependency of the sum of eigenvalues. By contrast, we here evaluate the temperature dependency of the sum of singular values. Instead of the quantity $S$ in Eq.~\eqref{area}, we evaluate the following quantity:
\begin{align}
\label{eq_area_svd}
S^{\mathrm{svd}} = \sum^{R'}_{r=1}\left(\frac{1}{2}\left(w^{\mathrm{svd}}_{r+1} + w^{\mathrm{svd}}_{r}\right)\left(T_{r+1}-T_{r}\right)\right),
\end{align}
where $w^{\mathrm{svd}}_{r}$ is the sum of singular values at the $r$-th temperature.

\begin{figure}[tb]
\centering
\includegraphics[width=75mm]{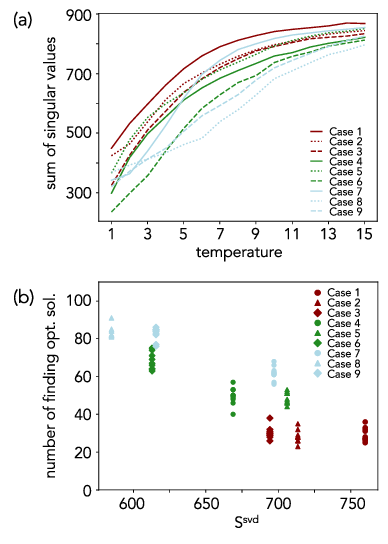}
\caption{(Color online) Numerical results of the SVD for randomly generated problems. (a) shows the temperature-dependency of the total singular values. (b) shows the relationship between the number of finding times of the optimal solution in 100 trials and the feature $S^{\mathrm{svd}}$.}
\label{fig_random_svd}
\end{figure}

\begin{figure}[tb]
  \centering
  \includegraphics[width=75mm]{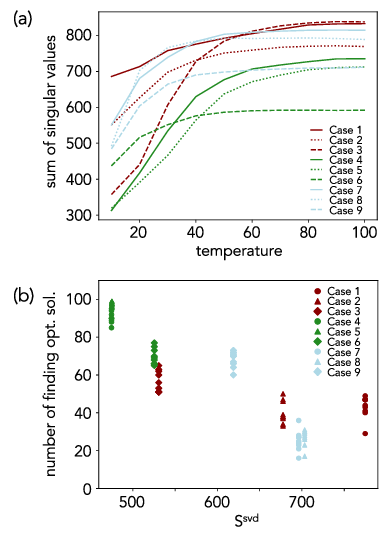}
  \caption{(Color online) Numerical results of the SVD for the 0-1 quadratic knapsack problems. (a) shows the temperature-dependency of the total singular values. (b) shows the relationship between the number of finding times of the optimal solution in 100 trials and the feature $S^{\mathrm{svd}}$.}
  \label{fig_qkp_svd}
\end{figure}

Figure~\ref{fig_random_svd}(a) shows the temperature-dependency of the sum of singular values. The values of $R'$ are 15 for all cases. Figure~\ref{fig_random_svd}(b) shows the relationship between the number of finding times of the optimal solution in 100 trials and the feature $S^{\mathrm{svd}}$. Figures~\ref{fig_qkp_svd}(a) and (b) show the corresponding results for the 0-1 quadratic knapsack problems; here, $R'$ in Eq.~\eqref{eq_area_svd} are as follows:
\begin{itemize}
\item $R'=5$ for Case 3, 4, 6, 7, 8, and 9,
\item $R'=6$ for Case 2 and 5,
\item $R'=8$ for Case 1.
\end{itemize}
The observation from Figs.~\ref{fig_qkp_svd}(a) and (b) are the same as the random problem cases.

From the comparison of the results of the EDMD and SVD analyses, we obtain the following observations:
\begin{itemize}
\item The EDMD and SVD analyses show negative correlations between the proposed features and difficulty in problem-solving.
\item The sums of Koopman eigenvalues take small values at low temperatures in all cases, but those of singular values take various values at low temperatures.
\item While Figs.~\ref{fig_random_svd}(a) and \ref{fig_qkp_svd}(a) show slow temperature-dependency, the EDMD results show a steep temperature dependency, as shown in Figs.~\ref{fig_random_Koopman}(a) and \ref{fig_qkp_Koopman}(a). Hence, a narrower temperature range is sufficient for the EDMD analysis than the SVD, contributing to lower computational costs.
\end{itemize}

We consider that the steeper temperature dependency in the Koopman analysis stems from the fact that the EDMD represents the dynamics itself. That is, the SVD focuses on a large matrix of the entire time series data. By contrast, the EDMD requires snapshot pairs, and then we focus only on transitions. As the temperature increases, the dynamics should change, and the EDMD could immediately reflect the details of the dynamics because of the analysis of the snapshot pairs. Therefore, the change in dynamics with a small temperature change would be apparent in the EDMD analysis.

These numerical experiments confirm that the data analysis flow based on the Koopman analysis works well in discussing the difficulty in problem-solving.

We finally comment on the difference between the values of the singular values and those of Koopman eigenvalues. The above numerical results indicate that the sum of singular values is larger than that of the Koopman eigenvalues. We consider the reason as follows. Since the SVD uses entire time series data, it analyzes a large matrix. Note that the SVD yields orthogonal matrices, and the norm of each column or row vector in the orthogonal matrix is $1$. Hence, A large matrix gives a long vector, and each component in the long vector should be small to set the norm to $1$. Hence, the singular values tend to take large values to reproduce the original matrix. By contrast, the Koopman analysis focuses on snapshot pairs, so the situation is different from the SVD. As a result, the sum of singular values is generally larger than that of the Koopman eigenvalues.

\section{Conclusion}
\label{sec_conclusion}

In this paper, we clarified the relationship between the Koopman analysis and the difficulty in problem-solving for combinatorial optimization problems. Since only a single time-series data at a temperature is not enough to discuss the difficulty, we generated multiple time-series data at various temperatures by using the replica exchange Monte Carlo method; as denoted above, a type of specialized hardware has employed the replica exchange Monte Carlo method for the parallel computation. In the data analysis flow, we proposed the quantity that aggregates the information of the multiple time-series data. The numerical experiments confirmed the validity of the proposals; there is a negative correlation between the proposed quantity and the ability of the solution search.

This paper is the first work to use Koopman analysis to discuss the difficulty in problem-solving for combinatorial optimization problems. There are some remaining tasks left for the future. For example, one should check the algorithm dependency of the analytical results; although the Metropolis rule employed in this paper is popular in optimization, the time-series data depends on the update algorithm. It would also be interesting to evaluate the algorithm with a mechanism to get out of the local solution, for example, as proposed in Ref.~\cite{Sato2024}. One of the other remaining tasks is the reduction of the computational costs in the EDMD; for example, the online EDMD \cite{Zhang2019} could be effective for sequential data acquisition in the replica exchange Monte Carlo method. In addition, investigation of the correspondence between the proposed and physical quantities is a subject for future work. The Koooman eigenvalue reflects the complexity of the snapshot pairs, not of the entire time-series data, and we confirmed that the simple summation of the Koopman eigenvalue corresponds to an indicator of difficulty. However, the quantity $S$ could have a connection to some physical quantity.

There are many recent studies for specialized hardware for combinatorial optimization problems, and customizing the algorithm for each problem would be beneficial. We believe that this paper will lead to future studies to quantify the difficulty in problem-solving.

\begin{acknowledgments}
This work was supported by JST FOREST Program (Grant Number JPMJFR216K, Japan).
\end{acknowledgments}

\end{document}